\documentclass[oncolumn]{aastex631}

\newcommand\h     {$^{\rm h}$}
\newcommand\m     {$^{\rm m}$}
\newcommand\s     {$^{\rm s}$}

\newcommand{\kms} {\ifmmode{\rm \,km\,s^{-1}}\else\,km\,s$^{-1}$\xspace\fi}

\newcommand{\ha}{\hbox{H$\alpha$}}
\newcommand{\hb}{\hbox{H$\beta$}}

\newcommand{\sii}{\hbox{[S\,{\sc ii}]}}
\newcommand{\nii}{\hbox{[N\,{\sc ii}]}}
\newcommand{\oiii}{\hbox{[O\,{\sc iii}]}}

\newcommand{\cgcg}{CGCG 020-010}

\shortauthors{Chung et al.}

\begin{document}

\title{Witnessing a Transformation to Blue-cored Dwarf Early-type Galaxies in Filaments and the Cluster Outskirts: Gas-phase Abundances and Internal Kinematics Perspectives}

\correspondingauthor{Jiwon Chung}
\email{jiwon@kasi.re.kr}

\author[0000-0003-0469-345X]{Jiwon Chung}
\affiliation{Korea Astronomy and Space Science Institute}
\author[0000-0003-3451-0925]{Joon Hyeop Lee}
\affiliation{Korea Astronomy and Space Science Institute}
\author[0000-0002-0145-9556]{Hyunjin Jeong}
\affiliation{Korea Astronomy and Space Science Institute}
\author[0000-0003-3474-9047]{Suk Kim}
\affiliation{Research Institute of Natural Sciences, Chungnam National University, Daejeon 34134, Republic of Korea}

\received{March 3, 2023}
\revised{March 21, 2023}
\accepted{April 4, 2023}
\begin{abstract}
The presence of transitional dwarf galaxies in filaments and cluster outskirts may be closely related to pre-processing in the filament; however, the underlying mechanism is not yet comprehensively understood. We present the spatially resolved chemical and kinematical properties of three blue-cored dwarf early-type galaxies (dE(bc)s) in the Virgo cluster and Virgo-related filaments (Crater $\&$ Virgo III) using the Sydney-AAO Multi-object Integral-field spectrograph (SAMI) Galaxy Survey. We map the spatial distribution of $\ha$, oxygen abundance (O/H), nitrogen-to-oxygen abundance ratio (N/O), stellar population age, and gas-stellar internal kinematics. We find irregular shapes of enhanced star-forming regions from the centers to the outlying regions of blue cores in dE(bc)s. These regions are relatively metal-poor compared with the surrounding regions, rendering the overall metallicity gradient of each galaxy positive. Furthermore, they exhibit higher N/O ratios at a given O/H relative to their surroundings, implying metal-poor gas infall by external processes. The equivalent width of the $\ha$ emission line in metal-poor regions indicates young age of star formation, 6-8 Myr. The disturbed ionized gas velocity field, one of the most prominent features of galaxy mergers is also discovered in two dE(bc)s. We propose that a moderately dense filament environment is favorable for the formation of blue cores in dEs, in which dE(bc)s in filaments may have already been transformed before they fall into the Virgo cluster. This process may contribute to the composition of galaxy population at the outskirts of the cluster.

\end{abstract}



\section{Introduction} \label{sec:intro}

In the standard theory of large-scale structure formation, the currently observable cosmic web is formed through the gravitational enhancement of small dark matter density fluctuations present in the early Universe \citep{deLapparent1986}. The large-scale structure of the Universe is characterized by the presence of filaments that intersect at nodes wherein clusters of galaxies are found \citep{Bond1996}. Hierarchical models of structure formation predict that a cluster of galaxies grows by the continuous accretion of galaxies from filaments. Presumably, 40 percent of the mass in the Universe resides within filaments \citep{Aragon-Calvo2010}, which causes them to be a non-negligible population from the perspective of galaxy evolution.

To date, various efforts have been dedicated to understand the role of filament environments based on galaxy properties. Some studies have reported that the filament environment is responsible for enhanced star formation activity and increased fraction of star-forming galaxies \citep{Porter2007,Fadda2008,Biviano2011,Coppin2012,Darvish2014}. On the other hand, galaxies in filaments exhibit low star formation rates, red colors, and low {\hbox{H\,{\sc i}}} mass fractions, thereby being pre-processed \citep{Martinez2016,Kuutma2017,Mahajan2018,Bonjean2020,Castignani2021,Lee2021}. 

\citet{Chung2021} investigated the chemical properties of star-forming dwarf galaxies (SFGDs) in the field, the Virgo cluster, and five Virgo-related filaments. They found an environmental disparity between filaments, in which SFDGs in a specific filament exhibit lower star formation rates and higher metallicities than those in the field, which is similar to the Virgo cluster counterparts. Meanwhile, SFDGs in the other four filaments exhibit field-like properties, indicating that each filament can have different environments even though they are connected to the same cluster. Becasue of the environmental disparity between Virgo filaments, in-depth understanding of galaxies in filaments and beyond to cluster outskirts in terms of the evolutionary pathway of galaxies in the large-scale structure of the Universe is necessary.

Dwarf early-type galaxies (dEs) containing dwarf ellipticals and dwarf lenticulars are the numerically predominant galaxy type, particularly in clusters and groups; dEs are more numerous than any other type of galaxy and are scarce in low-density fields \citep{Sandage85,Binggeli88}. Despite their ordinary and superficial appearances, the complexity and diversity of their characteristics have become evident in terms of kinematics \citep{Toloba2015,Chung2019}, structural features \citep{lisker2006a,lisker2006b,lisker2007}, and gaseous components \citep{Gu2006,Chung2021}. Thus, most dEs are considered to be transformed from late-type progenitors in the cluster environments because they ought to be sensitive to their surroundings and less able to retain their gaseous contents in the face of external perturbation due to their shallow gravitational potential wells.

In this respect, blue-cored dEs, dubbed dE(bc)s---those on the way to transforming red dEs from late-type progenitors---showing blue colors at their center due to recent or ongoing star formation \citep{derijcke2003,cellone05,Gu2006,lisker2006b,tully2008,derijcke2013,pak2014} are ideal laboratories for examining the mechanisms that govern galaxy transformation by environmental effects. For the dE(bc)s in the cluster, the ram pressure stripping has been recognized to be responsible for the transformation of dwarf galaxies, in which the interstellar medium of the outerpart of dwarf galaxies is stripped by interaction with a hot intracluster medium in the cluster \citep{lisker2006b}. As a plausible scenario, \citet{Smith2010} suggested that the late-type dwarf could be transformed to an early-type morphology by high-speed encounters with massive companions in a high-density cluster environment, which triggers central star formation with a short-lived excess of gas density. 

Notably, dE(bc)s are found in moderately dense regions such as outskirts of clusters, groups, and filaments \citep{tully2008,pak2014,Chung2021}. \citet{Kuutma2017} also found the trend that morphology transformation of a galaxy from late-type to early-type is more dominant in the filament than low-density void and suggested that this trend could possibly be related to a galaxy mergers or cut-off of gas supplies in near and inside filaments. 

For invaluable information on galaxy evolution,gas-phase metallicity could be used as a unique observable parameter for probing the chemical evolution of galaxies. As galaxies evolve, nucleosynthesis releases metals into the surrounding interstellar medium during the later evolutionary stages of the stars. As the cycle of star formation recurs, the metals in the interstellar medium of a galaxy become steadily enriched in the closed-box case. Thus, the metallicity traces the fossil record of the star formation history. The star formation history is linked to the gas interchange with the surroundings; therefore, the gas-phase metallicity becomes sensitive to the surrounding environment. 

The internal kinematics of galaxies also provides essential information for discriminating various mechanisms related to the formation of the blue-core in dE. The memories of the processes that occur during galaxy evolution are recorded in internal kinematics. In particular, the angular momentum of the stellar population of a galaxy is conserved even when the ram pressure of a cluster strips the interstellar medium of a galaxy \citep{Boselli2008a,Boselli2008b,Benson2015}. Moreover, the merging events frequently leave kinematically decoupled stellar structures to the remnant galaxies, which act as important clues for tracing such assembly histories \citep{Toloba2014}.

In this study, we investigate the formation of dE(bc)s in the context of the role of environments, using gas-phase chemical properties and internal kinematics of dwarf galaxies based on integral field spectroscopy (IFS) data. The IFS data provide powerful insights into the evolution of galaxies through the spatial distribution of gas-phase abundances and internal kinematics across objects. This paper is organized as follows:  In Section 2, we describe the data and analysis of spectra based on IFS data of third and final data release of the Sydney-AAO Multi-object Integral-field spectrograph (SAMI) Galaxy Survey \citep{Croom2021}. In Section 3, we explore the $\ha$, oxygen abundance, nitrogen-to-oxygen abundance, stellar population age, and internal kinematics at the spatially resolved scale. We also discuss possible formation scenarios from a pre-processing perspectives. Finally, we summarized our main results in Section 4. \\



\begin{figure}
\epsscale{1.2}
\plotone{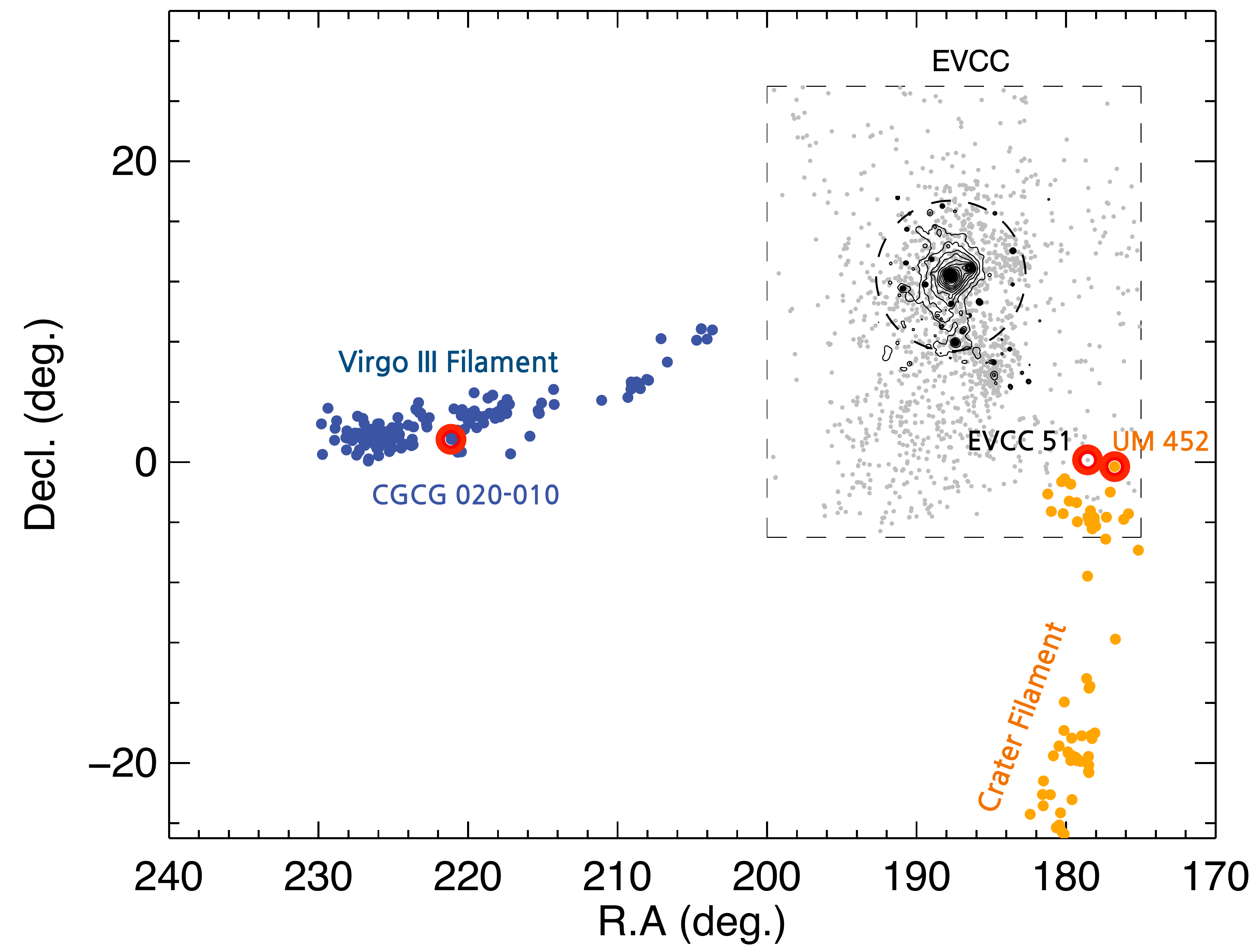} 

\caption{Projected spatial distribution of galaxies in the Virgo-related filaments (filled circles with different colors), the Virgo cluster (gray dots within the dashed box). Three dE(bc)s are overplotted with red open circles. The large dashed box is the region of the EVCC (Kim et al. 2014).  Contours represent the X-ray diffuse emission distribution of the cluster from ROSAT. The large dashed circle indicates the virial radius of Virgo A (5$\degr$.37 = 1.55 Mpc) defined by Ferrarese et al. (2012), considering a distance of 16.5 Mpc to the Virgo cluster from us (Jerjen et al. 2004; Mei et al. 2007). }
\label{spatial}
\end{figure} 

\section{DATA and ANALYSIS} \label{sec:intro}
\subsection{Data}
The SAMI is a multi-object IFS at the prime focus of the 3.9m Anglo Australian Telescope (AAT) and provides a 1 degree diameter field-of-view (FOV). The SAMI uses 13 fused fiber bundles (Hexabundles; \citealt{Bland-Hawthorn2011,Bryant2014}) with a filling factor of 75$\%$. Each bundle includes 61 fibers of 1.6 arcsec diameter to create an IFU with a 15 arcsec of diameter. The SAMI uses 580V grating at 3700-5700\AA\ in the blue arm and 1000R grating at 6250–7350\AA\ in the red arm, affording spectral resolutions of R=1812 ($\sigma$=70 km s$^{-1}$) and R=4263 ($\sigma$=30 km s$^{-1}$), respectively.

\subsection{Sample Selection}
The visually classified dE(bc)s sample in this study is based on the  Sloan Digital Sky Survey (SDSS) Data Release 12 (DR12; \citealt{Alam15}) galaxy sample reported by \citet{Chung2021}, which includes dwarf galaxies in the Extended Virgo Cluster Catalog (EVCC; \citealt{Kim2014}) and Virgo-related filamentary structures \citep{Kim2016}. These dE(bc)s were cross-matched with the SAMI IFS data. As a result, three dE(bc)s UM 452 and \cgcg\, and EVCC 51 were detected in the SAMI survey.

Figure ~\ref{spatial} shows the projected distribution of galaxies in the two Virgo-related filaments and the Virgo cluster. The three dE(bc)s are indicated by open red circles. The UM 452 and \cgcg\ reside in the Crater and the Virgo III filaments, which stretch to the south and east of the Virgo cluster, respectively. The EVCC 51 is located at the outskirts of the Virgo cluster with projected distances of 4.40 Mpc ($\sim$ 2.84 virial radius) from the Virgo cluster center, considering the distance of the Virgo cluster and virial radius of the Virgo A (1.55 Mpc; \citealt{Ferrarese2012}) and has a radial velocity of 1131 km s$^{-1}$. Thus, EVCC 51 is defined as a galaxy located inside a spherical symmetric infall model in a plot of radial velocity versus the cluster-centric distance of galaxies (see \citealt{Kim2014} for more details). The basic characteristics of our three dE(bc)s are summarized in Table ~\ref{table1}. 

 EVCC 51 is defined as galaxy located inside a spherical symmetric infall model

\begin{table*}[ht]
\caption{Summary of Sample Galaxies} 
\centering 
\begin{tabular}{l l l l l l l} 
\hline\hline 
Name  & SAMI ID & R.A. (deg.)    & Decl. (deg.)  & cz (km s$^{-1}$) & M$_i$ (mag.) & Environment   \\  

\hline 

UM 452  & 54102 & 176.753 & -0.294  & 1468 & -17.64 & Crater Filament\\
CGCG 020-010 & 240202 & 221.128 & 1.522 & 1473 & -16.96 & Virgo III Filament\\
EVCC 51  & 70114 & 178.551 & 0.137 & 1131 & -17.82 & Virgo Cluster  \\

\hline 
\end{tabular}
\label{table1}

\end{table*}


\subsection{Measurement of Emission Line Flux and Kinematics}

Our analysis is based on the third and final release of SAMI data cubes with adaptive binning, which uses the Voronoi method \citep{Cappellari2003} to obtain a median blue-arm signal-to-noise ratio (S/N) of 10 at least for each bin. For the bright regions at the central parts of each galaxy, one spaxel corresponds to a bin, whereas up to 43, 21, and 4 spaxels are summed into a bin in the faint outskirts of the UM 452, CGCG 020-010, and EVCC 51, respectively. The total number of bins for UM 452, \cgcg, and EVCC 51 are 243, 337, and 650, respectively. We utilized the Penalized Pixel cross-correlation Fitting ($\it pPXF$) algorithm of \citet{Cappellari2004} to model the stellar continuum in each bin simultaneously using stellar templates from the MILES library \citep{Vazdekis2010} and to estimate the kinematics of the stellar and gas components separately. For the kinematics, we apply pPXF to the observed spectrum with masking only bad pixels, and subsequently, pPXF yields the stellar, Balmer, and forbidden line kinematics. For the emission line fluxes, however, we measured them using the following processes:
(1) Apply pPXF to the observed spectrum with masking not only bad pixels but also all known emission lines ($\pm$500 km s$^{-1}$ range of each line). (2) The stellar continuum and absorption lines were determined via pPXF. (3) The residual spectrum was obtained by subtracting the continuum + absorption lines from the observed spectrum. (4) From the residual spectrum, the flux of each line was measured by Gaussian fitting. .

For correction of internal extinction to the measured emission line fluxes, we used the Balmer decrement of H$\alpha$/H$\beta$=2.86 and H$\gamma$/H$\beta$=0.47 for Case B emissivity with a temperature of 10,000 K and an electron density of 100 cm$^{-3}$ \citep{Osterbrock2006}. For all bins in each galaxy, the lowest S/N ratio is $\sim$ 12 for the weakest emission line $\nii\lambda6584$ used in this study.

\subsection{Emission Line Diagnostics}
\begin{figure}
\epsscale{1.0}
\plotone{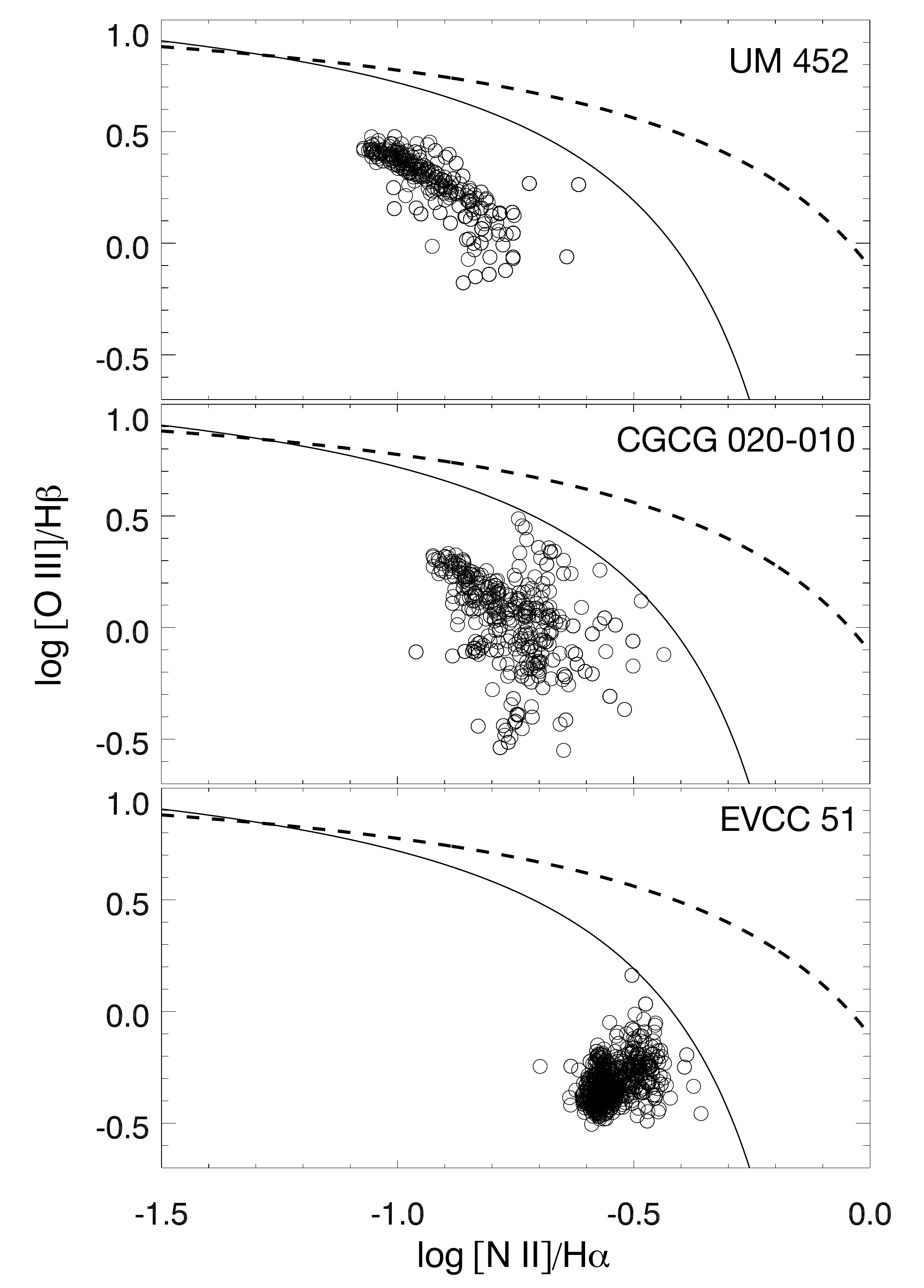} 

\caption{$\oiii\lambda5007/\hb$ vs. $\nii\lambda6584/\ha$ diagnostic diagram of three dE(bc)s, UM 452 in the Crater Filament (top), \cgcg\ in the Virgo III filament (middle), and the EVCC 51 in the Virgo cluster (bottom). The circles represent all bins dominated star formation processes that lie below the empirical curve of \citet{Kauffmann2003} denoted by the solid curve. The dashed curve presents the theoretical maximum starburst model line of \citet{Kewley2001}. }
\label{bpt}
\end{figure}

\citet{Baldwin81} proposed a diagnostic diagram to classify the dominant energy sources in emission-line galaxies. This diagram is commonly known as the Baldwin–Phillips–Terlevich (BPT) diagram and is used to classify galaxies according to the excitation mechanism of their emission lines. The BPT diagram consists of excitation-dependent and extinction-independent line ratios such as log($\oiii\lambda5007$/$\hb$) and log($\nii\lambda6584$/$\ha$). \citet{Kewley2001} used a photoionization model to formulate a theoretical maximum starburst line in a BPT diagram. The critical starburst line is determined by the upper limit of the theoretical stellar photoionization models. \citet{Kauffmann2003} modified the scheme proposed by \citep{Kewley2001} to rule out the possible AGN-{\hbox{H\,{\sc ii}}} composite galaxies. Figure ~\ref{bpt} shows BPT diagrams for all bins in three dE(bc)s in the Crater (top) and Virgo III (middle) filaments, as well as the Virgo cluster (bottom). All bins belonged to the star-forming criteria, implying that their interstellar medium was ionized by young and hot stellar populations. The EVCC 51, with a  lower $\oiii\lambda5007$/$\hb$ ratio, indicates that the majority of the star-forming regions are less excited than the other two dE(bc)s.


\subsection{Measurement of Gas-phase Abundance}
The oxygen abundance (O/H), the most abundant metal in the interstellar medium of a galaxy, is represented as a proxy of the metallicity of a galaxy. Various methods for oxygen abundance estimation have been proposed based on electron temperature ($T$$_e$). The ratio of the $\oiii\lambda4363$ to $\oiii\lambda\lambda4959,5007$ allows us to directly evaluate the oxygen abundance through the $T$$_e$ of the ionized gas. Unfavorably, these requisite lines are intrinsically faint, particularly in the low excited high metallicity regions of a galaxies. Therefore, we alternatively rely on a so-called empirical calibration to estimate the the metallicity in the ionized gas. The N2 (log($\nii\lambda6584$/$\ha$)) and O3N2 (($\oiii\lambda5007$/$\hb$)/($\nii\lambda6584$/$\ha$)) parameters have been calibrated against oxygen abundances derived directly by determining the $T$$_e$ \citep{Denicolo2002, Marino2013}. We estimated the metallicity for each bin in galaxies using the N2 metallicity indicator \citep{Denicolo2002} because N2 is less sensitive to dust extinction than O3N2. This N2 calibration is based on the relationship between the $T$$_e$ metallicities and the N2 using a local sample of galaxies, expressing the metallicity as 
\begin{equation}
  12+log(O/H) = 9.12+N2\times0.73.
\end{equation}
We note that we also examined the metallicity map using O3N2 calibration and it showed results consistent with the metallicity map using N2 calibration.

We estimated the nitrogen-to-oxygen (N/O) abundance ratio using the following empirical relationship derived by  \citep{Perez2009}: 
\begin{equation}
  log(N/O) = 1.26\times N2S2-0.86,
\end{equation}
where N2S2 is the N/O abundance ratio indicator, N2S2 = log($\nii\lambda6584$/($\sii\lambda\lambda6717,6731$).

The average uncertainties of O/H and N/O are $\sim$0.02 dex and $\sim$0.05 dex for the UM 452 and the \cgcg\ and $\sim$0.02 dex and $\sim$0.06 dex for the EVCC 51. The uncertainties of the O/H and N/O were derived from the flux errors of the emission lines, and did not include the systematic uncertainties in the N2 and N2S2 calibration.

\



\begin{figure*}
\epsscale{1.0}
\plotone{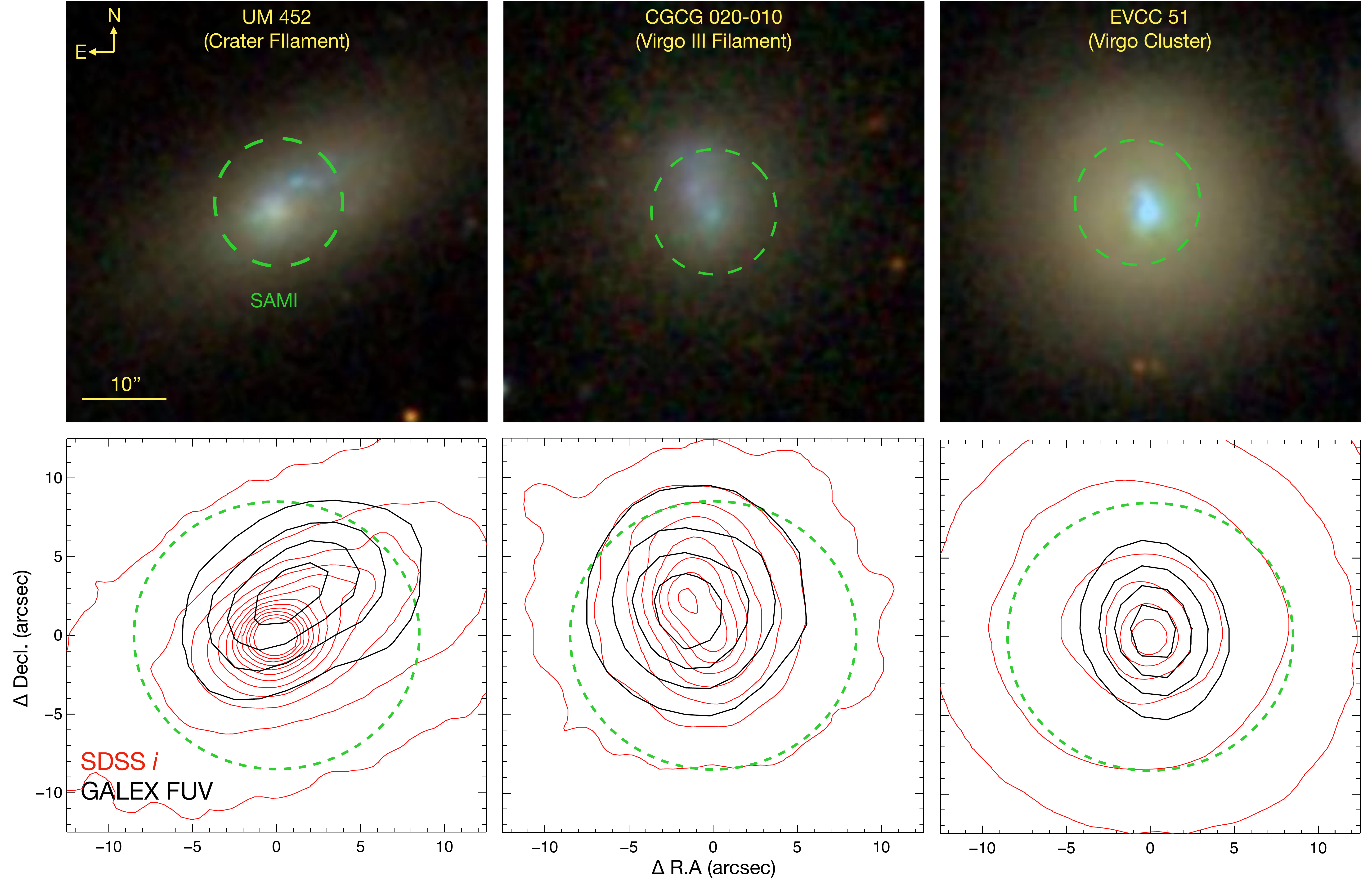} 

\caption{Top: SDSS $g$, $r$, and $i$ composite images of the UM 452 (left), the \cgcg\ (middle), and the EVCC 51 (right), respectively. Bottom: Isophote contours of SDSS $i$-band (red contours) and {\it GALEX} FUV band (black contours), respectively. The green dashed circles represent the circular FOV of the SAMI hexabundle. The lowest contour levels for $i$-band and FUV band were chosen at the about 3$\sigma$ level of the local background. Every subsequent contour level is increased by twice the value of the lowest level.
}
\label{sdss}
\end{figure*} 

\section{Results and Discussion}

\subsection{Photometric Characteristics}

Figure ~\ref{sdss} shows SDSS $g$, $r$, and $i$ composite images of three dE(bc)s and their flux contours of $i$-band and far-ultraviolet (FUV) band of Galaxy Evolution Explorer ({\it GALEX}; \citealt{martin2005}). The dashed circle and box represent FOV of the hexabundle and the cube of SAMI, respectively. The physical sizes of UM 452, \cgcg, and EVCC 51 in the FOV of SAMI hexabundle are about 1.5 kpc, 1.5 kpc, and 1.2 kpc, respectively. The $\ha$ emission traces the present star formation activity from young massive stars with timescales of 10 Myr, while the FUV flux traces less massive stars with timescales of a few hundred Myr \citep{Kennicutt1998}. 

These dE(bc)s exhibit the presence of blue cores in their central regions with respect to the outer parts of a galaxies. This is in agreement with the {\it GALEX} FUV contours, indicating the existence of young stellar populations and distinct star formation history in their center. UM 452 displays an off-centered FUV contour distribution compared with the $i$-band contours. When it comes to the star formation history, it indicates a hint that the recent star formation occurred in the region away (north-west direction) from the center of a galaxy where the old stellar population is dominant. Unlike the UM 452, the other two dE(bc)s show that their $i$-band and FUV-band contours are not significantly different.


\subsection{$\ha$ Emission Line Map}

\begin{figure}
\epsscale{0.5}
\plotone{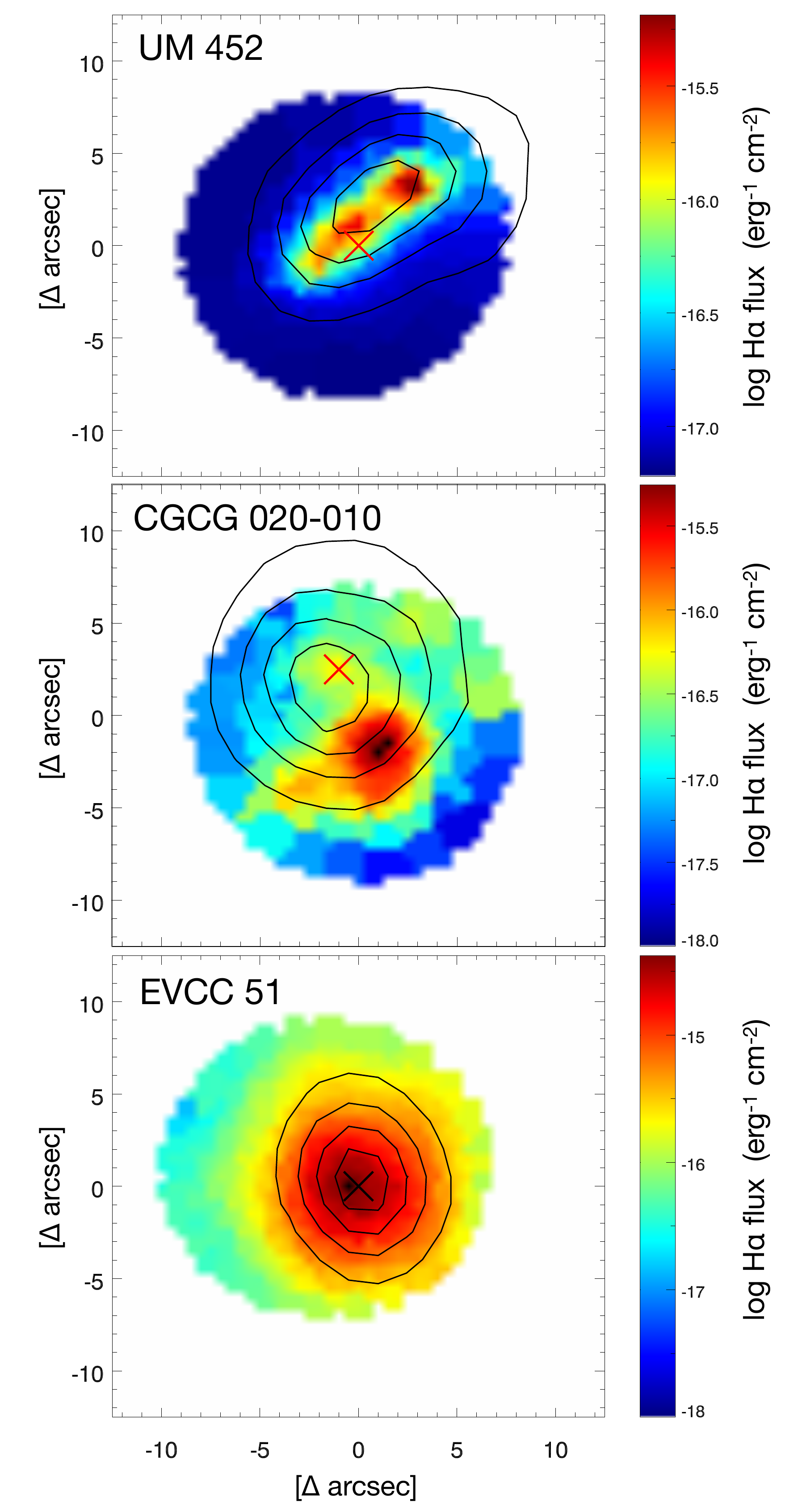} 

\caption{$\ha$ maps of the UM 452 (top), \cgcg\ (middle), and EVCC 51 (bottom), respectively. The cross indicates the center of SDSS $i$-band contours. The solid contours represent the distribution of {\it GALEX} FUV band. The color of each bin shows $\ha$ flux scaled as the color bar on the right-hand side. }
\label{ha}
\end{figure}

The $\ha$ emission line represents a region of ionized gas, and thus represents a young stellar population ($<$ 10 Myr), symbolizing a recent star-forming history. In Figure ~\ref{ha}, we present the $\ha$ emission line maps of three dE(bc)s in each environment. As the most notable feature, the UM 452 and the \cgcg\ include elongated star-forming regions that deviates from galaxy centers dominated by the old stellar population, which corresponds to the peak of the $i$-band luminosity (black contours). For the UM452, it can be seen that the elongated star-forming region is in good agreement with the FUV distribution stretching from the southeast to the northwest. This implies constant occurrence of star formation in this region from an age of at least 100 Myr. It is also noteworthy that UM 452 also has an elongated {\hbox{H\,{\sc i}}} structure along the southeast-northwest axis, which is well consistent with the elongated shape of an active star-forming region (see \citealt{Taylor1995} for details). Moreover, UM 452 is a gas-rich system compared with other elliptical galaxies \citep{Salzer2000}. In the case of {\cgcg}, the $\ha$ emission lines are distributed in a tail-like structure, slightly deviating from the centers of distribution of the FUV with the $i$-band, indicating the occurrence of star formation in this region at least 10 Myr ago. These off-centered star-forming clumps are expected in a scenario of gas infall onto the disk by merging with a galaxy \citep{Ceverino2016,Chung2019,Egorova2021}. Meanwhile, the star formation of EVCC 51 is mainly concentrated in the central part of a galaxy and coincides with the centers in the $i$-band and FUV-band, respectively.


\subsection{Gas-phase Abundances}

\subsubsection{Oxygen Abundance}


\begin{figure}
\epsscale{0.5}
\plotone{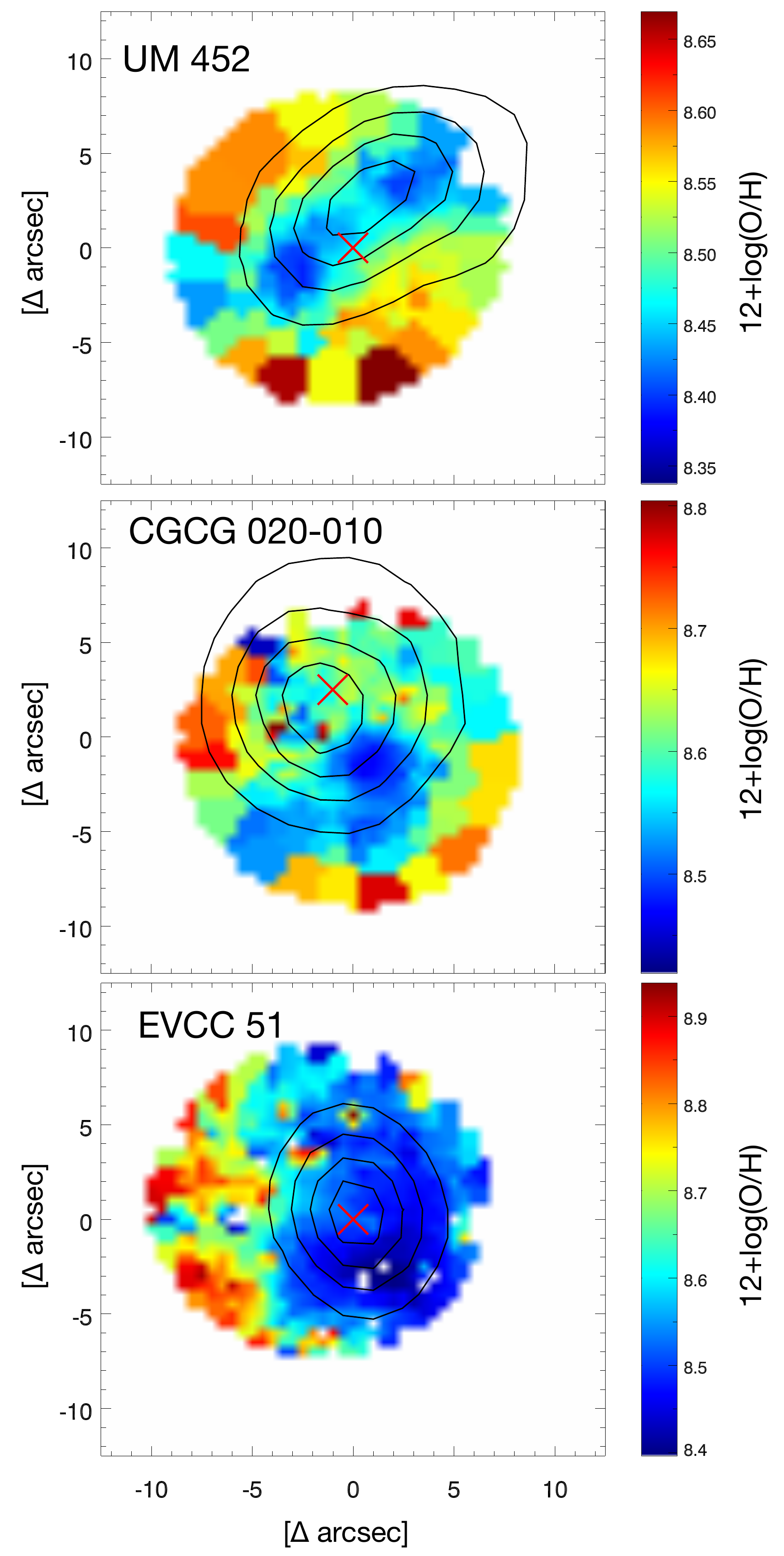} 

\caption{O/H maps of the UM 452 (top), \cgcg\ (middle), and EVCC 51 (bottom), respectively. The cross and solid contours are same as in Figure ~\ref{ha}. The color of each bin shows O/H scaled as the color bar on the right-hand side.}
\label{oxy}
\end{figure} 

The most abundant metallicity content, oxygen is produced by short-lived young massive stars, and its abundance reflects the star formation history of a galaxy by gas inflow and outflow. From the perspective of galaxy evolution by gas interchange, O/H may be an important parameter. Figure ~\ref{oxy} shows the distribution of O/H for our three dE(bc)s. Regarding the distribution of O/H as a function of radius, one remarkable result is that our three dE(bc)s tend to have lower metallicities in the enhanced star formation regions, suggesting that strong interactions or merging is the primary triggering agent of their star formation activities (see also Figure ~\ref{ha}). In general, metallicity distributions of low-mass star-forming galaxies are homogeneous, whereas high-mass star-forming galaxies have a negative metallicity gradient beacuase of the inside-out growth \citep{Kobulnicky1997,Belfiore2017}. According to the inside-out growth model, the metallicity gradient should have a negative slope because star formation mainly occurs in the central part of a galaxies prior to the star formation outskirts of a galaxy. For a low-mass galaxy, the fraction of gas transformed into stars at the center of a galaxy could be attenuated by galactic wind owing the shallow potential well, resulting in a flat metallicity gradient \citep{Brook12}. 

On the other hand, in line with previous results regarding the positive metallicity gradient, the lower O/H at the center of the galaxies supports the idea that the metal-poor region is a result of pristine gas infall and is associated with the galaxy merger \citep{Rupke2008,Rupke2010,Peeples2009,Ekta2010,Kewley2010,Chung2019}, which is consistent with our results. Some simulation studies about galaxy merging suggest that gas in the outer regions of both disks is rapidly driven into the central region of a galaxy during the merging, and then the intense central star formation is triggered \citep{Barnes96,DiMatteo07}. Besides, the inflow of metal-poor pristine gas is responsible for metallicity dilution in the central region \citep{Montuori10, Bustamante18}. In this sense, the relatively metal-poor regions showing enhanced star formation activities for our three dE(bc)s can be explained by a scenario in which the inflow of metal-poor gas causes dilution in the already metal-enriched regions.

\subsubsection{Nitrogen to Oxygen Abundance Ratio}


\begin{figure*}
\epsscale{1.2}
\plotone{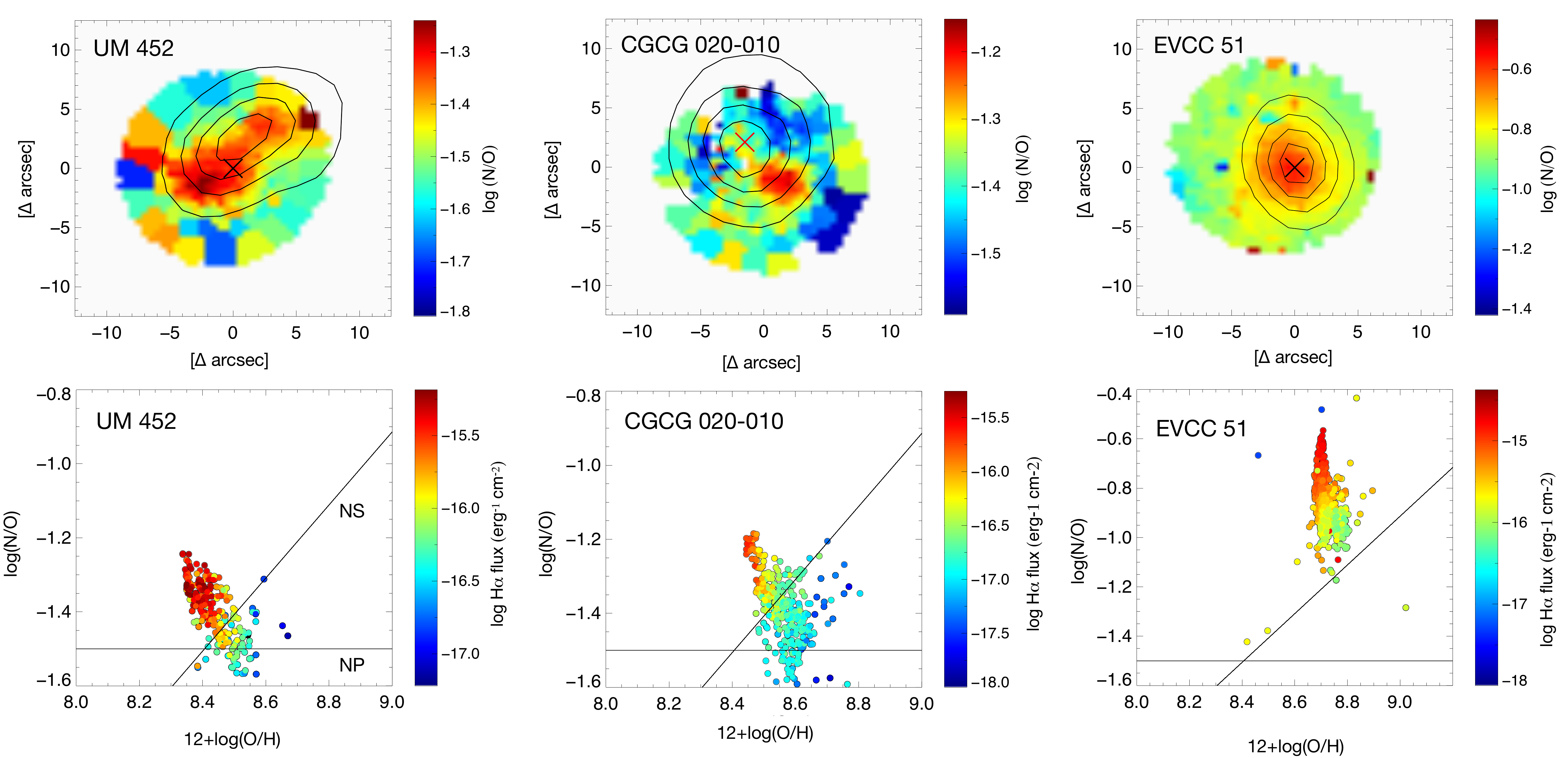} 

\caption{Top: /O ratios maps of the UM 452 (left), \cgcg\ (middle), and EVCC 51 (right), respectively. The cross and solid contours are the same as in Figure ~\ref{ha}. The color of each bin shows N/O scaled as the color bar on the right-hand side. Bottom: The N/O--O/H diagram for each dE(bc). The color of each bin shows $\ha$ flux scaled as the color bar on the right-hand side. The solid lines indicate the expectations for the primary (NP) and secondary (NS) origins of nitrogen taken from \citet{Vila-Costas1993}.}
\label{nitro}
\end{figure*} 

It is generally believed that the main contributors of nitrogen are the short-lived young massive stars and long-lived intermediate-mass stars that are progenitors of asymptotic giant branch stars. The nitrogen-to-oxygen abundance ratio (N/O) is constant for the low-metallicity regime (12+log(O/H) $\lesssim$ 8.4) by primary nitrogen origin in massive stars, whereas the both primary and secondary nitrogen contributions from intermediate-mass stars produces a positive correlation between N/O and O/H in the metal-rich regime \citep{Izotov1999,pilyugin03}. Therefore, the N/O abundance ratio as a function of the O/H ratio is an essential constraint on the chemical evolution of galaxies by gas inflow and the contributions of stellar populations \citep{Edmunds1978, Torres1989, Vila-Costas1993, Izotov2006, Chung2013}.

Figure ~\ref{nitro} shows the N/O distributions (top panels) and N/O versus O/H (bottom panels) of our three dE(bc)s. The colors of the circle symbols are coded according to the $\ha$ flux (a proxy for star formation activity) of the star-forming regions in the dE(bc)s. In the top panels of Figure ~\ref{nitro}, higher N/O ratios in the three dE(bc)s tend to be exhibited in relatively metal-poor regions where $\ha$ fluxes were enhanced (see also Figures ~\ref{ha} and ~\ref{oxy}). These trends may lead to intriguing distribution in the N/O--O/H diagram. The bottom panels of Figure ~\ref{nitro} show distributions of star-forming regions of three dE(bc)s in the N/O--O/H diagrams, which deviate from the general trend that the N/O ratio increases with increasing O/H in the regime of 12+log(O/H) $>$ 8.4 \citep{pilyugin03}. As mentioned above, metal-poor gas infall by merger might explain the simultaneously high N/O and relatively low O/H ratios of dE(bc)s due to the horizontal offset toward lower O/H at given N/O values in the N/O--O/H diagram \citep{Koppen2005, Amorin2010}. Moreover, this horizontal offset is more pronounced in central active star-forming regions (see color-coded circles) where metal-poor gas is expected to accumulate by the galaxy merger \citep{Montuori10}.


\begin{figure*}
\epsscale{1.2}
\plotone{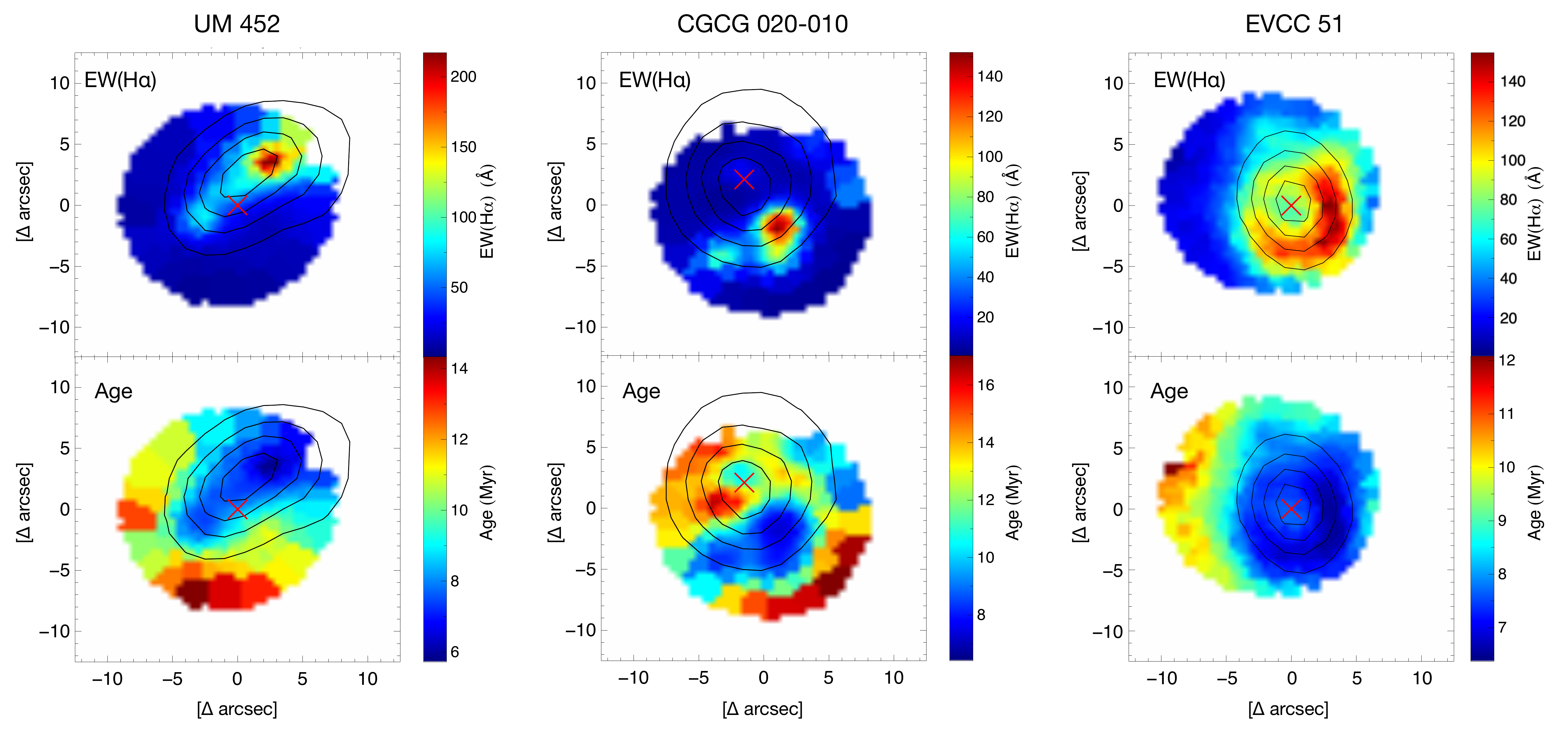} 

\caption{Map of EW($\ha$) (upper panels) and age (lower panels) of UM 452 (left), \cgcg\ (middle), and EVCC 51 (right), respectively.  Age map in Myr calculated from Starburst99 models at the constant metallicity of Z = Z$_{\odot}$ (mean of metallicities of the three dE(bc)s). The color of each bin shows EW($\ha$), an age scaled as the color bar on the right-hand side. Cross and solid contours are the same as those in Figure ~\ref{ha}.    }

\label{age}
\end{figure*} 
In addition to metal-poor gas inflow from mergers, Wolf-Rayet (WR) stars contributing to nitrogen production can also be an additional contributor to the increase in N/O at a given O/H \citep {Pustilnik2004, Chung2013}. As galaxies undergo mergers, they provide subsequent intense starbursts, increasing the relative fraction of massive young stars. Therefore, high N/O ratios at a given O/H may be associated with an additional contribution of WR star as well as metal-poor gas infall by merger. From the perspective of N/O enhancement at a given O/H, we checked whether the dE(bc)s have prominent WR features of blue-bump containing emission lines such as {\hbox{[N\,{\sc iii}]}}$\lambda$4640, {\hbox{[C\,{\sc iii}]}}$\lambda$4652, and {\hbox{[He\,{\sc ii}]}}$\lambda$4686. However, in our data, we could not find any WR feature around the {\hbox{[He\,{\sc ii}]}}$\lambda$4686 emission line. Nevertheless, we should keep in mind that detecting weak WR features in the spectrum can be deterred by the quality of the spectrum. As an example corresponding to this case, EVCC 51 is already classified as a WR galaxy with blue-bump detection around the {\hbox{[He\,{\sc ii}]}} emission line \citep{Zhang2007,Brinchmann2008}. Therefore, we cannot completely rule out the possibility that the other two dE(bc)s, UM 452 and \cgcg\, also include WR stars. Thus, we propose that an efficient gas infall is likely to prevail in the central regions of dE(bc)s, which would be responsible for the metallicity dilution with N/O enhancement by recent star formation.

\subsubsection{Age of Stellar Population} 

In the top panels of Figure ~\ref{age}, we first plot the maps of the equivalent width (EW) of the strongest Balmer lines $\ha$ for each dE(bc) to estimate the age of the stellar populations contributing to relatively metal-poor star-forming regions that appear to be undergoing a gas infall. For UM 452 and EVCC 51, the EW($\ha$) distributions show different features compared with the $\ha$ flux map, as shown in Figure ~\ref{ha}. The UM 452 has EW($\ha$) peak in the northwest direction from the center, and the EVCC 51 exhibits a curvilinear structure in the southwest direction.

We compare the EW($\ha$) with the predictions of the evolutionary synthesis models of Starburst99 \citep{Leitherer1999} at constant mean metallicities of three dE(bc)s (Z = Z$_{\odot}$). The UM 425, \cgcg\, and EVCC 51 have mean metallicities of 8.50, 8.62, and 8.71, respectively. This is similar to the solar value, assuming 12+log(O/H)$_{\odot}$ = 8.69 \citep{Allende-Prieto2001}. For generating models, we choose Geneva tracks with standard mass-loss rates, along with assumptions of an instantaneous star formation at time-steps of 0.1 Myr with a Salpeter initial mass function. We also adopted the expanding atmosphere of Padrach/Hillier. The bottom panels of Figure ~\ref{age} represent the age maps of the stellar population for three dE(bc)s in the range of about 6-17 Myr. For our three dE(bc)s, the irregular shapes of star-forming regions near the centers of the $i$-band contour appear to have formed almost coevally approximately 6-8 Myr ago, under the assumption of an instantaneous star formation, whereas their surroundings appear slightly older. In the case of EVCC 51, this result also corresponds to the result of \citet{Koleva2013}, wherein the central region of EVCC 51 is mainly composed of an old stellar population but lacks an intermediate-age population, and recently, a single young starburst occurred by gas infall or episodic star formation.


\begin{figure*}
\epsscale{1.2}
\plotone{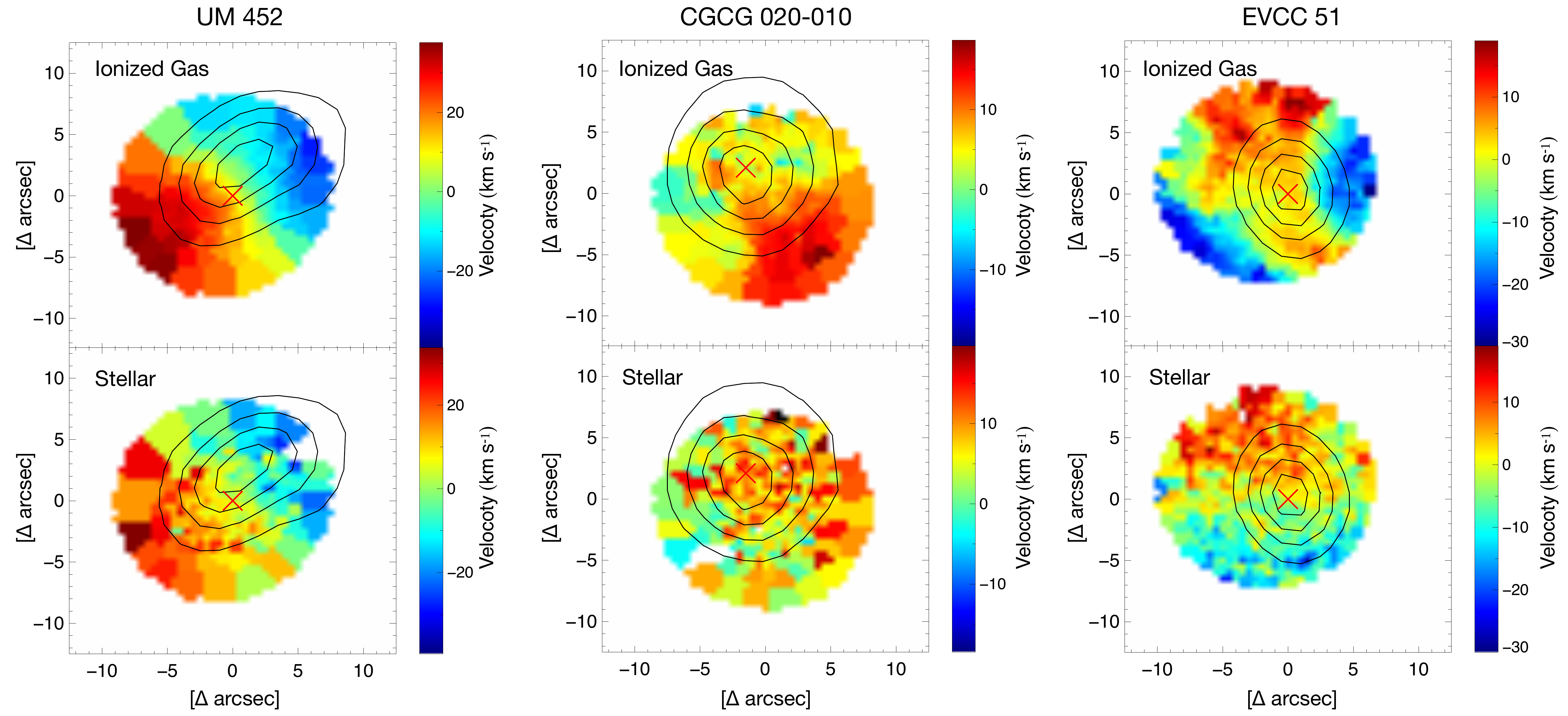} 

\caption{Kinematic map of ionized gas (upper panels) and stellar (lower panels) of the UM 452 (left), the \cgcg\ (middle), and the EVCC 51 (right), respectively. The Cross and solid contours are the same as those in Figure ~\ref{ha}.}
\label{kine}
\end{figure*} 

\subsection{Role of Filament Concerning Evolutionary path of Dwarf Galaxies} 

Under the framework of a galaxy population in large-scale structures of the Universe, pre-processing in dwarf galaxies has significant implications for the assembly history of cluster galaxies in relation to their star formation histories. The morphologies of galaxies are related to the physical conditions of their environment in which galaxies are located \citep{Dressler1980,Binggeli1987}. Interestingly, dE(bc)s are abundantly found in moderately dense regions, such as the infall regions of clusters, groups, and filaments. It has been reported that the fractions of dE(bc)s in the filaments (up to $\sim$24$\%$ of all dwarfs; \citealt{Chung2021}) and groups ($\sim$70$\%$ of dEs; \citealt{tully2008, pak2014}) are higher than those in the Virgo cluster ($\sim$5$\%$ of dEs; \citealt{lisker2006a, lisker2007,Meyer2014}). This allows us to conclude that the majority of dE(bc)s are not transformed by ram pressure stripping and high-speed multiple encounters with massive galaxies in cluster environments.

In contrast to the cluster environments, mergers and interactions between galaxies are more frequent in low-density environments with relatively low-velocity dispersions \citep{Binney1987, Boselli2006}  and can effectively contribute to the transformation of galaxies \citep{Vijayaraghavan2013}. The infall region, which is at the junction between a filament and a galaxy cluster, extends from the outskirts of a cluster up to several virial radii. \citet{Chung2019} found the kinematically decoupled features in two dE(bc)s located in the infall region of the Virgo cluster, indicating the possibility that these may have undergone gas-rich dwarf-dwarf mergers (pre-processing) before they subsequently fall into the cluster. Furthermore, \citet{Chung2021} have also found that a large amount of dE(bc)s lies in a group-like environment in the Virgo III filament to which \cgcg\ belongs and proposed that tidal interactions between galaxies in the filaments could be the probable formation mechanism of dE(bc)s. The EVCC 51 located in the infall region of the cluster outskirts is close to the boundary of the Crater filament, and the projected distance between EVCC 51 and UM 452 is 0.53 Mpc (see Figure ~\ref{spatial}) with a relative velocity of $\sim$ 340km s$^{-1}$. Assuming that galaxy merging/interaction is a formation channel for three dE(bc)s found in filaments and cluster outskirts, this event would have occurred in moderately dense environments in filaments before they subsequently fell into the cluster.

Figure ~\ref{kine} shows the maps of ionized gas based on the Balmer lines (top panels) and stellar kinematics (bottom) for UM 452, \cgcg\ (in the Virgo-connected filaments) and EVCC 51 (in the Virgo cluster). In the case of UM 452, the velocity maps show rotating ionized gas and stellar components along the major axis (SE-NW) of a galaxy. The \cgcg\ does not show prominent stellar rotation, whereas ionized gas in the southeast region is redshifted. The non-rotating feature of \cgcg\ is explained by the merger scenario. In particular, the mass ratio of the merger is the dominant factor in which an equal-mass merger is more favorable for the formation of slow or non-rotating systems than a high mass ratio one \citep{Jesseit2009}. Similarly, the collision angle of the merging orbit and gas fraction also affects to merger remnants \citep{Jesseit2009, Choi2017}. In lower-mass regimes, nonrotating dwarf spheroidal galaxies can also be formed from mergers between disky dwarf galaxies \citep{Valcke2008, Kazantzidis2011, Ebrova2015}. Moreover, the \cgcg\ shows the more redshifted ionized gas velocity in the SW region, while stellar components are slow- or non-rotating. The EVCC 51 also shows that stellar components clearly rotate from north to south, whereas ionized gas components both in NE and SW are redshifted. These kinematical properties of \cgcg\ and EVCC 51 can be explained by a scenario wherein galaxy mergers are responsible for disturbed kinematics owing to external processes that lead to disruption of the velocity field  \citep{Shapiro2008,Bellocchi2012}.

Furthermore, there are also similarities between the remnant of gas-rich dwarf-dwarf mergers and dE(bc)s. The numerical simulations have confirmed that the gas-rich dwarf–dwarf merging can trigger central star formation through the gas infall, forming a massive compact blue core dominated by young stellar populations \citep{Bekki2008, Watts2016}. At the late stage of a merger, the merged galaxy may appear to be as a blue compact dwarf (BCD) galaxy with an elliptical shape as a starburst galaxy, and subsequently becoming a galaxy morphologically similar to a dE(bc) after the last BCD phase. Such BCDs showed a positive age gradient in the numerical simulation of \citet{Bekki2008}, and this is comparable to our observations that the {\hbox{H\,{\sc ii}}} regions in the central parts of the three target dE(bc)s have lower O/H and younger ages than the outer part of galaxies (see Figure ~\ref{age}). Recent studies based on the IFS data also found a hint of gas infall in a BCD \citep{Ju2022}. Furthermore, inconsistency between N/O and O/H in the enhanced star formation region of a BCD \citep{Kumari2017,Kumari2018} originated from metallicity dilution by mergers. Based on long-slit spectroscopy, \cite{Chung2019} have also found that dE(bc)s in the infall region of the Virgo cluster exhibit similar structural properties in scaling relations. Furthermore, high central circular velocity gradients were also found as a result of the dwarf-dwarf merger.

Another plausible scenario for the formation of dE(bc)s is the accretion of cold gas from the intrafilament medium that replenishes filament galaxies with pristine {\hbox{H\,{\sc i}}} gas \citep{Keres2005, Sancisi2008, Darvish2015, Kleiner2017}. \citet{Ristea2022} have also proposed that gas accretion from the filament is the main driver of producing the disturbed velocity field. With a similar notion, quiescent phase of BCDs during repeated episodic star formation might be observed as transitional dwarf galaxies such as dE(bc)s and eventually evolve into dEs \citep{Sanchez-Almeida2008}.


\section {Summary and Conclusion}

In this paper, we presented the spatially resolved properties of gas-phase abundances and ionized and stellar kinematics in two dE(bc)s in filaments physically connected to the Virgo cluster and one dE(bc) in the Virgo cluster using SAMI IFS data. The main results are as follows:

1. In terms of $\ha$ view, two out of three dE(bc)s have irregularly shaped active star-forming regions that slightly deviate from a galaxy centers dominated by the old stellar population, which corresponds to the peak of the i-band luminosity. These off-centered star-forming clumps indicate the hint of gas infall.

2. Our three dE(bc)s exhibit distinctly metal-poor regions (positive metallicity gradients) in the central region of dE(bc)s, whereas dwarf galaxies usually have flat-metallicity gradients. Interestingly, the distributions of star-forming regions in the N/O-O/H diagram show an unusual trend, in which more excited star-forming regions exhibit enhanced N/O ratios at a given O/H. This indicates that strong interaction or merging may be the primary triggering agent of metallicity dilution and star formation activity by metal-poor gas infall.

3. We compared the observed EW(\ha) with the prediction of evolutionary synthesis models of Starburst99 to estimate the age of the stellar population in the relatively metal-poor regions. The age of the blue core in the dE(bc)s is approximately 6-8Myr, whereas their surroundings appear slightly older.

4. The disturbed velocity field, one of the most prominent features of a galaxy merger, was discovered in the \cgcg\ and EVCC 51. Therefore, we speculate that the dwarf-dwarf merger may be an important mechanism of the dE(bc) formation.

The discovery of a large fraction of dE(bc) in the filaments and evidence of gas infall reveal the transformation mechanism of dwarf galaxies wherein the majority of dE(bc)s may not have formed in cluster environments. We uncovered evidence of gas infall via chemical and kinematical properties in filaments and the infall region of the Virgo cluster from their spatially resolved properties based on IFS data. Consequently, we can conclude that a moderately dense filament environment is favorable for the formation of blue cores in dEs. This implies that dE(bc)s may have already been preprocessed in the filament before they fell into the Virgo cluster. This process may contribute to the composition of galaxy population at the outskirts of the cluster. In the future, coupled with the IFS data of a larger sample of dE(bc)s in various environments, we intend to enhance our understanding of the formation mechanisms of blue core in early-type dwarf galaxies. 

\

We are grateful to the anonymous referee for helpful comments and suggestions that improved the clarity and quality of this paper. J.C. acknowledges support from the Basic Science Research Program through the National Research Foundation (NRF) of Korea (2018R1A6A3A01013232, 2022R1F1A1072874). J.H.L. $\&$ H.J. acknowledge support from the Basic Science Research Program through the National Research Foundation (NRF) of Korea (2022R1A2C1004025, $\&$ 2019R1F1A1041086, respectively). S.K. acknowledges support from the Basic Science Research Program through the National Research Foundation (NRF) of Korea(2019R1I1A1A01061237 $\&$ 2022R1C1C2005539). This work was supported by the Korea Astronomy and Space Science Institute under the R$\&$D program (Project No. 2023-1-830-01) supervised by the Ministry of Science and ICT (MSIT). {\it GALEX} data presented in this paper were obtained from the Mikulski Archive for Space Telescopes (MAST) at the Space Telescope Science Institute. The specific observations analyzed can be accessed via doi: \dataset[10.17909/xv9g-ny36]{https://doi.org/10.17909/xv9g-ny36}.

\clearpage


\begin{thebibliography}{}
\bibliographystyle{aasjournal}
\bibitem[Alam et al.(2015)]{Alam15} Alam, S., Albareti, F.~D., Allende Prieto, C., et al.\ 2015, \apjs, 219, 12. doi:10.1088/0067-0049/219/1/12
\bibitem[Allende Prieto et al.(2001)]{Allende-Prieto2001} Allende Prieto, C., Lambert, D.~L., \& Asplund, M.\ 2001, \apjl, 556, L63. doi:10.1086/322874
\bibitem[Amor{\'\i}n et al.(2010)]{Amorin2010} Amor{\'\i}n, R.~O., P{\'e}rez-Montero, E., \& V{\'\i}lchez, J.~M.\ 2010, \apjl, 715, L128. doi:10.1088/2041-8205/715/2/L128
\bibitem[Arag{\'o}n-Calvo et al.(2010)]{Aragon-Calvo2010} Arag{\'o}n-Calvo, M.~A., van de Weygaert, R., \& Jones, B.~J.~T.\ 2010, \mnras, 408, 2163. doi:10.1111/j.1365-2966.2010.17263.x
\bibitem[Baldwin et al.(1981)]{Baldwin81} Baldwin, J.~A., Phillips, M.~M., \& Terlevich, R.\ 1981, \pasp, 93, 5
\bibitem[Barnes \& Hernquist(1996)]{Barnes96} Barnes, J.~E. \& Hernquist, L.\ 1996, \apj, 471, 115. doi:10.1086/177957
\bibitem[Bekki(2008)]{Bekki2008} Bekki, K.\ 2008, \mnras, 388, L10. doi:10.1111/j.1745-3933.2008.00489.x
\bibitem[Belfiore et al.(2017)]{Belfiore2017} Belfiore, F., Maiolino, R., Tremonti, C., et al.\ 2017, \mnras, 469, 151. doi:10.1093/mnras/stx789
\bibitem[Bellocchi et al.(2012)]{Bellocchi2012} Bellocchi, E., Arribas, S., \& Colina, L.\ 2012, \aap, 542, A54. doi:10.1051/0004-6361/201117894
\bibitem[Benson et al.(2015)]{Benson2015} Benson, A.~J., Toloba, E., Mayer, L., et al.\ 2015, \apj, 799, 171. doi:10.1088/0004-637X/799/2/171
\bibitem[Bland-Hawthorn et al.(2011)]{Bland-Hawthorn2011} Bland-Hawthorn, J., Bryant, J., Robertson, G., et al.\ 2011, Optics Express, 19, 2649. doi:10.1364/OE.19.002649
\bibitem[Bond et al.(1996)]{Bond1996} Bond, J.~R., Kofman, L., \& Pogosyan, D.\ 1996, \nat, 380, 603. doi:10.1038/380603a0
\bibitem[Bonjean et al.(2020)]{Bonjean2020} Bonjean, V., Aghanim, N., Douspis, M., et al.\ 2020, \aap, 638, A75. doi:10.1051/0004-6361/201937313
\bibitem[Binggeli et al.(1987)]{Binggeli1987} Binggeli, B., Tammann, G.~A., \& Sandage, A.\ 1987, \aj, 94, 251. doi:10.1086/114467
\bibitem[Binggeli et al.(1988)]{Binggeli88} Binggeli, B., Sandage, A., \& Tammann, G.~A.\ 1988, \araa, 26, 509. doi:10.1146/annurev.aa.26.090188.002453
\bibitem[Binney \& Tremaine(1987)]{Binney1987} Binney, J. \& Tremaine, S.\ 1987, Princeton, N.J. : Princeton University Press, c1987.
\bibitem[Biviano et al.(2011)]{Biviano2011} Biviano, A., Fadda, D., Durret, F., et al.\ 2011, \aap, 532, A77. doi:10.1051/0004-6361/201016174
\bibitem[Boselli et al.(2008a)]{Boselli2008a} Boselli, A., Boissier, S., Cortese, L., et al.\ 2008, \aap, 489, 1015. doi:10.1051/0004-6361:200809546
\bibitem[Boselli et al.(2008b)]{Boselli2008b} Boselli, A., Boissier, S., Cortese, L., et al.\ 2008, \apj, 674, 742. doi:10.1086/525513
\bibitem[Boselli \& Gavazzi(2006)]{Boselli2006} Boselli, A. \& Gavazzi, G.\ 2006, \pasp, 118, 517. doi:10.1086/500691
\bibitem[Brinchmann et al.(2008)]{Brinchmann2008} Brinchmann, J., Kunth, D., \& Durret, F.\ 2008, \aap, 485, 657. doi:10.1051/0004-6361:200809783
\bibitem[Brook et al.(2012)]{Brook12} Brook, C.~B., Stinson, G., Gibson, B.~K., et al.\ 2012, \mnras, 419, 771. doi:10.1111/j.1365-2966.2011.19740.x
\bibitem[Bryant et al.(2014)]{Bryant2014} Bryant, J.~J., Bland-Hawthorn, J., Fogarty, L.~M.~R., et al.\ 2014, \mnras, 438, 869. doi:10.1093/mnras/stt2254
\bibitem[Bustamante et al.(2018)]{Bustamante18} Bustamante, S., Sparre, M., Springel, V., et al.\ 2018, \mnras, 479, 3381. doi:10.1093/mnras/sty1692
\bibitem[Cappellari \& Copin(2003)]{Cappellari2003} Cappellari, M. \& Copin, Y.\ 2003, \mnras, 342, 345. doi:10.1046/j.1365-8711.2003.06541.x
\bibitem[Cappellari \& Emsellem(2004)]{Cappellari2004} Cappellari, M. \& Emsellem, E.\ 2004, \pasp, 116, 138. doi:10.1086/381875
\bibitem[Castignani et al.(2021)]{Castignani2021} Castignani, G., Combes, F., Jablonka, P., et al.\ 2021, arXiv:2101.04389
\bibitem[Cellone \& Buzzoni(2005)]{cellone05} Cellone, S.~A. \& Buzzoni, A.\ 2005, \mnras, 356, 41. doi:10.1111/j.1365-2966.2004.08422.x
\bibitem[Ceverino et al.(2016)]{Ceverino2016} Ceverino, D., S{\'a}nchez Almeida, J., Mu{\~n}oz Tu{\~n}{\'o}n, C., et al.\ 2016, \mnras, 457, 2605. doi:10.1093/mnras/stw064
\bibitem[Chung et al.(2021)]{Chung2021} Chung, J., Kim, S., Rey, S.-C., et al.\ 2021, \apj, 923, 235. doi:10.3847/1538-4357/ac3002
\bibitem[Chung et al.(2013)]{Chung2013} Chung, J., Rey, S.-C., Sung, E.-C., et al.\ 2013, \apjl, 767, L15. doi:10.1088/2041-8205/767/1/L15
\bibitem[Chung et al.(2019)]{Chung2019} Chung, J., Rey, S.-C., Sung, E.-C., et al.\ 2019, \apj, 879, 97. doi:10.3847/1538-4357/ab25e8
\bibitem[Choi \& Yi(2017)]{Choi2017} Choi, H. \& Yi, S.~K.\ 2017, \apj, 837, 68. doi:10.3847/1538-4357/aa5e4b
\bibitem[Coppin et al.(2012)]{Coppin2012} Coppin, K.~E.~K., Geach, J.~E., Webb, T.~M.~A., et al.\ 2012, \apjl, 749, L43. doi:10.1088/2041-8205/749/2/L43
\bibitem[Croom et al.(2021)]{Croom2021} Croom, S.~M., Owers, M.~S., Scott, N., et al.\ 2021, \mnras, 505, 991. doi:10.1093/mnras/stab229
\bibitem[Darvish et al.(2015)]{Darvish2015} Darvish, B., Mobasher, B., Sobral, D., et al.\ 2015, \apj, 814, 84. doi:10.1088/0004-637X/814/2/84
\bibitem[Darvish et al.(2014)]{Darvish2014} Darvish, B., Sobral, D., Mobasher, B., et al.\ 2014, \apj, 796, 51. doi:10.1088/0004-637X/796/1/51
\bibitem[de Lapparent et al.(1986)]{deLapparent1986} de Lapparent, V., Geller, M.~J., \& Huchra, J.~P.\ 1986, \apjl, 302, L1. doi:10.1086/184625
\bibitem[Denicol{\'o} et al.(2002)]{Denicolo2002} Denicol{\'o}, G., Terlevich, R., \& Terlevich, E.\ 2002, \mnras, 330, 69
\bibitem[De Rijcke et al.(2003)]{derijcke2003} De Rijcke, S., Zeilinger, W.~W., Dejonghe, H., et al.\ 2003, \mnras, 339, 225. doi:10.1046/j.1365-8711.2003.06171.x
\bibitem[De Rijcke et al.(2013)]{derijcke2013} De Rijcke, S., Buyle, P., \& Koleva, M.\ 2013, \apjl, 770, L26. doi:10.1088/2041-8205/770/2/L26
\bibitem[Di Matteo et al.(2007)]{DiMatteo07} Di Matteo, P., Combes, F., Melchior, A.-L., et al.\ 2007, \aap, 468, 61. doi:10.1051/0004-6361:20066959
\bibitem[Dressler(1980)]{Dressler1980} Dressler, A.\ 1980, \apj, 236, 351. doi:10.1086/157753
\bibitem[Ebrov{\'a} \& {\L}okas(2015)]{Ebrova2015} Ebrov{\'a}, I. \& {\L}okas, E.~L.\ 2015, \apj, 813, 10. doi:10.1088/0004-637X/813/1/10
\bibitem[Edmunds \& Pagel(1978)]{Edmunds1978} Edmunds, M.~G. \& Pagel, B.~E.~J.\ 1978, \mnras, 185, 77P. doi:10.1093/mnras/185.1.77P
\bibitem[Ekta \& Chengalur(2010)]{Ekta2010} Ekta, B. \& Chengalur, J.~N.\ 2010, \mnras, 406, 1238. doi:10.1111/j.1365-2966.2010.16756.x
\bibitem[Egorova et al.(2021)]{Egorova2021} Egorova, E.~S., Egorov, O.~V., Moiseev, A.~V., et al.\ 2021, \mnras, 504, 6179. doi:10.1093/mnras/stab1192
\bibitem[Fadda et al.(2008)]{Fadda2008} Fadda, D., Biviano, A., Marleau, F.~R., et al.\ 2008, \apjl, 672, L9. doi:10.1086/526457
\bibitem[Ferrarese et al.(2012)]{Ferrarese2012} Ferrarese, L., C{\^o}t{\'e}, P., Cuillandre, J.-C., et al.\ 2012, \apjs, 200, 4. doi:10.1088/0067-0049/200/1/4
\bibitem[Gu et al.(2006)]{Gu2006} Gu, Q., Zhao, Y., Shi, L., et al.\ 2006, \aj, 131, 806. doi:10.1086/498891
\bibitem[Izotov et al.(2006)]{Izotov2006} Izotov, Y.~I., Stasi{\'n}ska, G., Meynet, G., et al.\ 2006, \aap, 448, 955. doi:10.1051/0004-6361:20053763
\bibitem[Izotov \& Thuan(1999)]{Izotov1999} Izotov, Y.~I. \& Thuan, T.~X.\ 1999, \apj, 511, 639. doi:10.1086/306708
\bibitem[Jesseit et al.(2009)]{Jesseit2009} Jesseit, R., Cappellari, M., Naab, T., et al.\ 2009, \mnras, 397, 1202. doi:10.1111/j.1365-2966.2009.14984.x
\bibitem[Ju et al.(2022)]{Ju2022} Ju, M., Yin, J., Liu, R., et al.\ 2022, \apj, 938, 96. doi:10.3847/1538-4357/ac9056
\bibitem[Kauffmann et al.(2003)]{Kauffmann2003} Kauffmann, G., Heckman, T.~M., Tremonti, C., et al.\ 2003, \mnras, 346, 1055
\bibitem[Kazantzidis et al.(2011)]{Kazantzidis2011} Kazantzidis, S., {\L}okas, E.~L., Mayer, L., et al.\ 2011, \apjl, 740, L24. doi:10.1088/2041-8205/740/1/L24
\bibitem[Kennicutt(1998)]{Kennicutt1998} Kennicutt, R.~C.\ 1998, \araa, 36, 189
\bibitem[Kere{\v{s}} et al.(2005)]{Keres2005} Kere{\v{s}}, D., Katz, N., Weinberg, D.~H., et al.\ 2005, \mnras, 363, 2. doi:10.1111/j.1365-2966.2005.09451.x
\bibitem[Kewley et al.(2001)]{Kewley2001} Kewley, L.~J., Dopita, M.~A., Sutherland, R.~S., et al.\ 2001, \apj, 556, 121
\bibitem[Kewley et al.(2010)]{Kewley2010} Kewley, L.~J., Rupke, D., Zahid, H.~J., et al.\ 2010, \apjl, 721, L48. doi:10.1088/2041-8205/721/1/L48
\bibitem[Kim et al.(2016)]{Kim2016} Kim, S., Rey, S.-C., Bureau, M., et al.\ 2016, \apj, 833, 207
\bibitem[Kim et al.(2014)]{Kim2014} Kim, S., Rey, S.-C., Jerjen, H., et al.\ 2014, \apjs, 215, 22
\bibitem[Kleiner et al.(2017)]{Kleiner2017} Kleiner, D., Pimbblet, K.~A., Jones, D.~H., et al.\ 2017, \mnras, 466, 4692. doi:10.1093/mnras/stw3328
\bibitem[Kobulnicky \& Skillman(1997)]{Kobulnicky1997} Kobulnicky, H.~A. \& Skillman, E.~D.\ 1997, \apj, 489, 636. doi:10.1086/304830
\bibitem[Koleva et al.(2013)]{Koleva2013} Koleva, M., Bouchard, A., Prugniel, P., et al.\ 2013, \mnras, 428, 2949. doi:10.1093/mnras/sts238
\bibitem[K{\"o}ppen \& Hensler(2005)]{Koppen2005} K{\"o}ppen, J. \& Hensler, G.\ 2005, \aap, 434, 531. doi:10.1051/0004-6361:20042266
\bibitem[Kumari et al.(2017)]{Kumari2017} Kumari, N., James, B.~L., \& Irwin, M.~J.\ 2017, \mnras, 470, 4618. doi:10.1093/mnras/stx1414
\bibitem[Kumari et al.(2018)]{Kumari2018} Kumari, N., James, B.~L., Irwin, M.~J., et al.\ 2018, \mnras, 476, 3793. doi:10.1093/mnras/sty402
\bibitem[Kuutma et al.(2017)]{Kuutma2017} Kuutma, T., Tamm, A., \& Tempel, E.\ 2017, \aap, 600, L6. doi:10.1051/0004-6361/201730526
\bibitem[Lee et al.(2021)]{Lee2021} Lee, Y., Kim, S., Rey, S.-C., et al.\ 2021, \apj, 906, 68. doi:10.3847/1538-4357/abcaa0
\bibitem[Leitherer et al.(1999)]{Leitherer1999} Leitherer, C., Schaerer, D., Goldader, J.~D., et al.\ 1999, \apjs, 123, 3. doi:10.1086/313233
\bibitem[Lisker et al.(2006b)]{lisker2006b} Lisker, T., Glatt, K., Westera, P., et al.\ 2006, \aj, 132, 2432. doi:10.1086/508414
\bibitem[Lisker et al.(2006a)]{lisker2006a} Lisker, T., Grebel, E.~K., \& Binggeli, B.\ 2006, \aj, 132, 497. doi:10.1086/505045
\bibitem[Lisker et al.(2007)]{lisker2007} Lisker, T., Grebel, E.~K., Binggeli, B., et al.\ 2007, \apj, 660, 1186. doi:10.1086/513090
\bibitem[Mahajan et al.(2018)]{Mahajan2018} Mahajan, S., Singh, A., \& Shobhana, D.\ 2018, \mnras, 478, 4336. doi:10.1093/mnras/sty1370
\bibitem[Marino et al.(2013)]{Marino2013} Marino, R.~A., Rosales-Ortega, F.~F., S{\'a}nchez, S.~F., et al.\ 2013, \aap, 559, A114. doi:10.1051/0004-6361/201321956
\bibitem[Martin et al.(2005)]{martin2005} Martin, D.~C., Fanson, J., Schiminovich, D., et al.\ 2005, \apjl, 619, L1. doi:10.1086/426387
\bibitem[Mart{\'\i}nez et al.(2016)]{Martinez2016} Mart{\'\i}nez, H.~J., Muriel, H., \& Coenda, V.\ 2016, \mnras, 455, 127. doi:10.1093/mnras/stv2295
\bibitem[Meyer et al.(2014)]{Meyer2014} Meyer, H.~T., Lisker, T., Janz, J., et al.\ 2014, \aap, 562, A49. doi:10.1051/0004-6361/201220700
\bibitem[Montuori et al.(2010)]{Montuori10} Montuori, M., Di Matteo, P., Lehnert, M.~D., et al.\ 2010, \aap, 518, A56. doi:10.1051/0004-6361/201014304
\bibitem[Osterbrock \& Ferland(2006)]{Osterbrock2006} Osterbrock, D.~E. \& Ferland, G.~J.\ 2006, Astrophysics of gaseous nebulae and active galactic nuclei, 2nd. ed. by D.E. Osterbrock and G.J. Ferland. Sausalito, CA: University Science Books, 2006
\bibitem[Pak et al.(2014)]{pak2014} Pak, M., Rey, S.-C., Lisker, T., et al.\ 2014, \mnras, 445, 630. doi:10.1093/mnras/stu1722
\bibitem[Peeples et al.(2009)]{Peeples2009} Peeples, M.~S., Pogge, R.~W., \& Stanek, K.~Z.\ 2009, \apj, 695, 259. doi:10.1088/0004-637X/695/1/259
\bibitem[P{\'e}rez-Montero \& Contini(2009)]{Perez2009} P{\'e}rez-Montero, E. \& Contini, T.\ 2009, \mnras, 398, 949. doi:10.1111/j.1365-2966.2009.15145.x
\bibitem[Pilyugin et al.(2003)]{pilyugin03} Pilyugin, L.~S., Thuan, T.~X., \& V{\'\i}lchez, J.~M.\ 2003, \aap, 397, 487. doi:10.1051/0004-6361:20021458
\bibitem[Porter \& Raychaudhury(2007)]{Porter2007} Porter, S.~C. \& Raychaudhury, S.\ 2007, \mnras, 375, 1409. doi:10.1111/j.1365-2966.2006.11406.x
\bibitem[Pustilnik et al.(2004)]{Pustilnik2004} Pustilnik, S., Kniazev, A., Pramskij, A., et al.\ 2004, \aap, 419, 469. doi:10.1051/0004-6361:20035646
\bibitem[Ristea et al.(2022)]{Ristea2022} Ristea, A., Cortese, L., Fraser-McKelvie, A., et al.\ 2022, \mnras, 517, 2677. doi:10.1093/mnras/stac2839
\bibitem[Rupke et al.(2010)]{Rupke2010} Rupke, D.~S.~N., Kewley, L.~J., \& Chien, L.-H.\ 2010, \apj, 723, 1255. doi:10.1088/0004-637X/723/2/1255
\bibitem[Rupke et al.(2008)]{Rupke2008} Rupke, D.~S.~N., Veilleux, S., \& Baker, A.~J.\ 2008, \apj, 674, 172. doi:10.1086/522363
\bibitem[Salzer et al.(2000)]{Salzer2000} Salzer, J.~J., Gronwall, C., Lipovetsky, V.~A., et al.\ 2000, \aj, 120, 80. doi:10.1086/301418
\bibitem[S{\'a}nchez Almeida et al.(2008)]{Sanchez-Almeida2008} S{\'a}nchez Almeida, J., Mu{\~n}oz-Tu{\~n}{\'o}n, C., Amor{\'\i}n, R., et al.\ 2008, \apj, 685, 194. doi:10.1086/590380
\bibitem[Sancisi et al.(2008)]{Sancisi2008} Sancisi, R., Fraternali, F., Oosterloo, T., et al.\ 2008, \aapr, 15, 189. doi:10.1007/s00159-008-0010-0
\bibitem[Sandage et al.(1985)]{Sandage85} Sandage, A., Binggeli, B., \& Tammann, G.~A.\ 1985, \aj, 90, 1759. doi:10.1086/113875
\bibitem[Shapiro et al.(2008)]{Shapiro2008} Shapiro, K.~L., Genzel, R., F{\"o}rster Schreiber, N.~M., et al.\ 2008, \apj, 682, 231. doi:10.1086/587133
\bibitem[Smith et al.(2010)]{Smith2010} Smith, R., Davies, J.~I., \& Nelson, A.~H.\ 2010, \mnras, 405, 1723. doi:10.1111/j.1365-2966.2010.16545.x
\bibitem[Taylor et al.(1995)]{Taylor1995} Taylor, C.~L., Brinks, E., Grashuis, R.~M., et al.\ 1995, \apjs, 99, 427. doi:10.1086/192193
\bibitem[Toloba et al.(2015)]{Toloba2015} Toloba, E., Guhathakurta, P., Boselli, A., et al.\ 2015, \apj, 799, 172. doi:10.1088/0004-637X/799/2/172
\bibitem[Toloba et al.(2014)]{Toloba2014} Toloba, E., Guhathakurta, P., van de Ven, G., et al.\ 2014, \apj, 783, 120. doi:10.1088/0004-637X/783/2/120
\bibitem[Torres-Peimbert et al.(1989)]{Torres1989} Torres-Peimbert, S., Peimbert, M., \& Fierro, J.\ 1989, \apj, 345, 186. doi:10.1086/167894
\bibitem[Tully \& Trentham(2008)]{tully2008} Tully, R.~B. \& Trentham, N.\ 2008, \aj, 135, 1488. doi:10.1088/0004-6256/135/4/1488
\bibitem[Watts \& Bekki(2016)]{Watts2016} Watts, A. \& Bekki, K.\ 2016, \mnras, 462, 3314. doi:10.1093/mnras/stw1812
\bibitem[Valcke et al.(2008)]{Valcke2008} Valcke, S., de Rijcke, S., \& Dejonghe, H.\ 2008, \mnras, 389, 1111. doi:10.1111/j.1365-2966.2008.13654.x
\bibitem[Vazdekis et al.(2010)]{Vazdekis2010} Vazdekis, A., S{\'a}nchez-Bl{\'a}zquez, P., Falc{\'o}n-Barroso, J., et al.\ 2010, \mnras, 404, 1639. doi:10.1111/j.1365-2966.2010.16407.x
\bibitem[Vila-Costas \& Edmunds(1993)]{Vila-Costas1993} Vila-Costas, M.~B. \& Edmunds, M.~G.\ 1993, \mnras, 265, 199. doi:10.1093/mnras/265.1.199
\bibitem[Vijayaraghavan \& Ricker(2013)]{Vijayaraghavan2013} Vijayaraghavan, R. \& Ricker, P.~M.\ 2013, \mnras, 435, 2713. doi:10.1093/mnras/stt1485
\bibitem[Zhang et al.(2007)]{Zhang2007} Zhang, W., Kong, X., Li, C., et al.\ 2007, \apj, 655, 851. doi:10.1086/510231


\end{thebibliography}
\end{document}